# Scalable synthesis of WS$_2$ on graphene and h-BN: an all-2D platform for light-matter transduction

*Antonio Rossi,*[†‡] *Holger Büch*[†] *Carmine Di Rienzo,*[†‡] *Vaidotas Miseikis,*[†] *Domenica Convertino,*[†‡] *Ameer Al-Temimy,*[†] *Valerio Voliani,*[†] *Mauro Gemmi,*[†] *Vincenzo Piazza*[†] *and Camilla Coletti*[†§*]

[†]Center for Nanotechnology Innovation @NEST, Istituto Italiano di Tecnologia, Piazza S. Silvestro 12, 56127 Pisa, Italy

[‡]NEST-Scuola Normale Superiore, Piazza San Silvestro 12, 56127 Pisa, Italy

[§]Graphene Labs, Istituto Italiano di Tecnologia, Via Morego 30, 16163 Genova, Italy

*corresponding author: camilla.coletti@iit.it

**Abstract**

By exhibiting a measurable bandgap and exotic valley physics, atomically-thick tungsten disulfide (WS2) offers exciting prospects for optoelectronic applications. The synthesis of continuous WS2 films on other two-dimensional (2D) materials would greatly facilitate the implementation of novel all-2D photoactive devices. In this work we demonstrate the scalable growth of WS2 on graphene and hexagonal boron nitride (h-BN) via a chemical vapor deposition (CVD) approach. Spectroscopic and microscopic analysis reveal that the film is bilayer-thick, with local monolayer inclusions. Photoluminescence measurements show a remarkable conservation of polarization at room temperature peaking 74% for the entire WS2 film. Furthermore, we present a scalable bottom-up approach for the design of photoconductive and photoemitting patterns.

## 1. Introduction

Two-dimensional (2D) transition metal dichalcogenides (TMDs) have recently raised interest in the scientific community and beyond, due to their remarkable optoelectronic properties[1-3] and prospected integration into flexible and transparent platforms. In particular, tungsten disulfide

(WS$_2$) has been appreciated as the TMD of choice for photoactive devices. Indeed, atomically-thick WS$_2$ does not only feature a direct optical band gap of about 2 eV at the *K* and *K′* symmetry points of the Brillouin zone[4-6], but also exhibits long exciton life and coherence times, causing remarkable photoluminescence (PL)[4]. Most notably, as a direct consequence of giant spin-orbit coupling and spin-valley coupling, bilayer WS$_2$ presents an astonishing preservation of polarization at room temperature (RT)[7]. Up to now, polarization conservation has been achieved in conventional solid-state systems by adopting technologically unpractical cryogenic temperatures. Hence, WS$_2$ represents an exciting playground for light-matter interaction studies and a promising platform for valleytronic applications. To date, most of the remarkable properties of WS$_2$ have been observed by using mechanically exfoliated flakes. In particular, valley polarization has been measured for micrometer-sized WS$_2$ grains on SiO$_2$[7, 8]. The scalable synthesis of continuous WS$_2$ films displaying high polarization coherence is instrumental in the implementation of a WS$_2$-based valleytronic technology. Moreover, incorporation of such WS$_2$ films within van der Waals heterostacks would pave the way for the realization of novel all-2D optoelectronic devices. Chemical vapor deposition (CVD) is the most suitable technique for the scalable synthesis of highly-crystalline 2D heterostacks[9, 10]. However, such an approach is not trivial as weak interlayer interactions favor three-dimensional (3D) island growth[11]. The formation of multi-layer islands is typically avoided by adopting short reaction times which lead, however, to the synthesis of isolated crystals[11-14]. To date, the only works on direct growth of WS$_2$ on other 2D-materials have demonstrated isolated sub-micrometer grains of few-layer thick WS$_2$ on graphene[13] and isolated grains on hexagonal boron nitride (h-BN) substrates[14].

In this work we show that continuous atomic-thick WS$_2$ films can be synthesized on appealing 2D substrates, i.e. graphene and h-BN, by using CVD. Spectroscopic, microscopic and photoluminescence measurements indicate that on both substrates the synthesized WS$_2$ is bilayer-thick, with local monolayer inclusions. PL measurements reveal that the polarization of the incident light is highly retained. The polarization anisotropy retrieved from PL polarization maps over areas of thousands of square-microns peaks at 74% at room temperature. These results indicate the suitability of scalable CVD-grown $WS_2$ films to absorb and conserve quantum information in the form of polarized light and to open exciting prospects for the adoption of van der Waals heterostacks in valleytronics. Furthermore, we present a scalable bottom-up approach for the design of photoconductive and photoemitting patterns.

## 2. Methods

*2.1 WS$_2$ growth*

To synthesize WS$_2$, we used a vapor-phase reaction from solid sulfur (S) and tungsten trioxide (WO$_3$) powders in a horizontal quartz tube. The selected precursors have previously led to high quality 2D crystals of WS$_2$ with large domain sizes on classical bulk insulators[4, 15, 16]. Sulfurization of WO$_3$ powder (Sigma Aldrich, 99.995%) was carried out within a horizontal hot-wall furnace (Lenton PTF). The furnace comprises an inner hot zone, in which 10 mg of WO$_3$ were placed, and a cooler outer zone in which 1g of sulfur (S) (Sigma Aldrich, 99.998%) was placed (about ~10 cm away from the hot-zone edge). The growth temperature within the hot-zone was set to 900 °C. This final temperature was reached at a rate of 1°C/s and was kept for 1 hour and 30 minutes. The substrates were placed face-up next to WO$_3$ powder within the same crucible. Before the temperature ramp up, the chamber was pumped down to a pressure of ~5x10$^{-2}$ mbar. Argon was flown during the temperature ramp with a flux of 500 sccm, leading to a pressure of 4.5 mbar, which kept the sulfur solid. After reaching 900°C, the Ar flux was suddenly reduced to 80 sccm, which reduced the furnace pressure to 1.3 mbar thus initiating sulfur evaporation and WO$_3$ sulfurization. The combination of a sudden evaporation of S at the growth temperature of 900 °C and high reactant flow rates, were found to balance the competing processes of sublimation, reaction, transfer, diffusion and precipitation of the reactants in favor of fast and dense WS$_2$ growth. WS$_2$ synthesis was performed on h-BN flakes exfoliated on quartz, CVD graphene transferred on quartz, and epitaxial graphene on silicon carbide (SiC)[17, 18]. The choice of transparent substrates anticipates the use of the investigated heterostacks for optoelectronic applications.

*2.2 Photoluminescence microscopy*

Photoluminescence microscopy experiments were carried out with a Leica SP5 confocal laser scanning microscope using a 63x (NA 1.2) water-immersion objective. As light source, pulsed solid state laser diode at 640 nm was used (Picoquant). The luminescence arising from the sample was collected through the objective and filtered with a 640 notch filter and with a 610 long pass filter (Chroma). To measure luminescence polarization, a broadband (420-680) polarizer beam splitter was introduced in the light path. The two arising beams are collected with two identical fiber couple SPAPDs (Picoquant). The alignment of the polarizer beam splitter principal axis and the excitation laser polarization plane was proved by an additional internal polarizer filter to be better than 5°. See Supplementary Information for experimental details about substrates preparation, Raman analysis, transmission electron microscopy (TEM), scanning electron microscopy (SEM) and atomic force microscopy (AFM).

## 3 Results and discussion

*3.1 Strong light polarization conservation in $WS_2$ synthesized on h-BN*

We first analyze the properties of $WS_2$ synthesized on h-BN, an ideal substrate to preserve near-pristine properties of graphene[19] and other 2D materials. Figure 1(a) shows a SEM micrograph of a representative h-BN flake exfoliated on quartz and fully covered with $WS_2$. No isolated triangular crystal[11, 15, 20] is observed and the uniform distribution of $WS_2$ on the flake is confirmed by Raman spectroscopy, which is instrumental in identifying the nature and the thickness of TMDs[21, 22]. The intensity ratio of the in-plane vibrational mode $E_{2g}^1$ and of the out-of-plane vibrational mode $A_{1g}$ is indicative of the thickness of $WS_2$ when using specific excitation wavelengths such as that adopted in this work, i.e., 532 nm[21, 23, 24] As visible in Fig. 1(b), the $A_{1g}/E_{2g}^1$ ratio is approaching 1, indicating bilayer coverage[23]. Interestingly, only along h-BN crystal anomalies (clearly visible in the optical image reported in the inset in panel (b)), a smaller ratio is observed, suggesting monolayer-thick regions. We further corroborate the monolayer/bilayer nature of the synthesized film by analyzing the overtone peak located at 310 cm$^{-1}$. This peak, which is attributed to rigid interlayer shear forces, diminishes with the number of layers, and ultimately disappears when approaching the monolayer limit[25, 26]. Indeed, as shown in Supporting Information, the 310 cm$^{-1}$ peak is barely detectable.

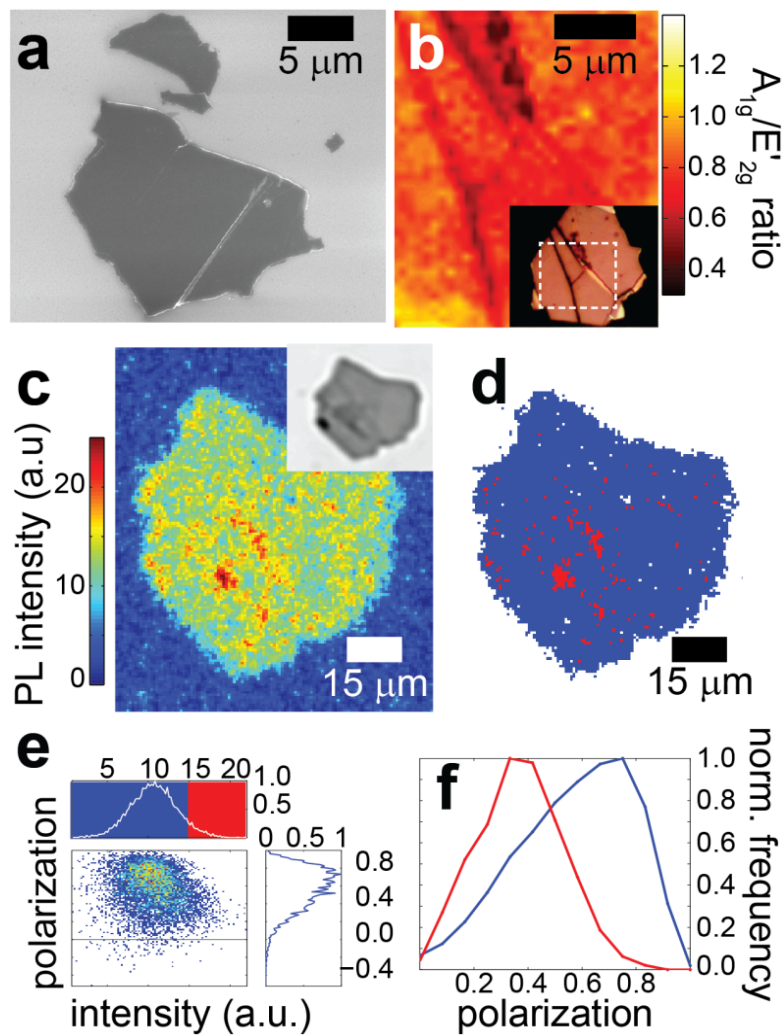

**Figure 1.** (a) SEM image of h-BN flakes exfoliated on quartz and fully covered by WS$_2$ crystals. (b) Raman map of the ratio between the intensities of A$_{1g}$ and E$_{2g}$$^1$ peaks taken in the area shown in the optical image in the inset. (c) Raw PL intensity map of a selected h-BN flake (also shown in the optical micrograph in the inset). (d) Binary PL intensity map, where the high PL intensity areas (in red) near the domain edges of the h-BN flakes are separated from the lower intensity values (in blue) elsewhere by defining a threshold, as shown in the intensity graph in (e). (e) 2D histogram of PL polarization vs. PL intensity with respective projections, revealing a correlation of the two quantities. (f) Histograms of the PL polarization within the high (red) and low (blue) PL intensity areas as shown in panels (d) and (e).

Time- and polarization-resolved PL spectroscopy is used to investigate the optical properties of the synthesized $WS_2$ film (see Experimental Section for details). This technique allows us to extract both the lifetime of the excitons created by the incoming photons and the conservation of their polarization. PL measurements were performed by using linearly polarized laser light with a near-resonant excitation energy (1.94 eV). To quantify the degree of polarization anisotropy of the emission, $P$, the following relation is used:

$$P = \frac{I_{\parallel} - I_{\perp}}{I_{\parallel} + I_{\perp}}, \qquad (1)$$

where $I_{\parallel}$ ($I_{\perp}$) is the intensity of PL with parallel (perpendicular) polarization with respect to the polarization of the excitation. As expected, PL emission is observed across the entire h-BN flake (see Figure 1(c)). A significantly stronger intensity – compatible with a transition to direct band-gap – is observed along h-BN crystal boundaries (compare map to inset in panel (c)), where monolayer $WS_2$ patches were indeed detected via Raman. In Figure 1 (e), we plot the PL intensity histogram (vertical graph), the PL polarization histogram (horizontal graph) and a 2D correlation histogram (intensity plot) obtained by analyzing the representative flake in panel (c). Remarkably, the peak polarization anisotropy is about 74% over an area of thousands of square micrometers. The elongated 2D histogram indicates a strong correlation between intensity and polarization, which we elaborate further by plotting separate histograms for areas attributed to monolayer $WS_2$ and bilayer $WS_2$ (red and blue in panel (e)), based on the PL intensity. The two histograms are shown in Figure 1(f) and define the polarization conservation to be 43% for monolayer $WS_2$ and 74% for bilayer $WS_2$, respectively. These values are in close agreement with those recently measured for isolated mono- and bilayer $WS_2$ crystals on $SiO_2$ (i.e., 30% and 80%) with circularly polarized light[8]. The slightly higher (lower) values measured for monolayer (bilayer) found in this work can be explained by the fact that both thicknesses occur below the resolution threshold of the PL setup (~ 250 nm), as also indicated by the wide tails of the histograms. Furthermore, the mean excitonic lifetime measured for the film via time-resolved PL is ~ 350 ps, which is in agreement with that reported in literature for bilayer $WS_2$[27]. It is to note that a continuous atomically-thick polycrystalline $WS_2$ film such as that presented in this work does not pose an obstacle for coherent photoabsorption/emission, since the exciton radius in $WS_2$ is as small as ~1-2 nm[28].

*3.2 Epitaxial growth of $WS_2$ on CVD graphene*

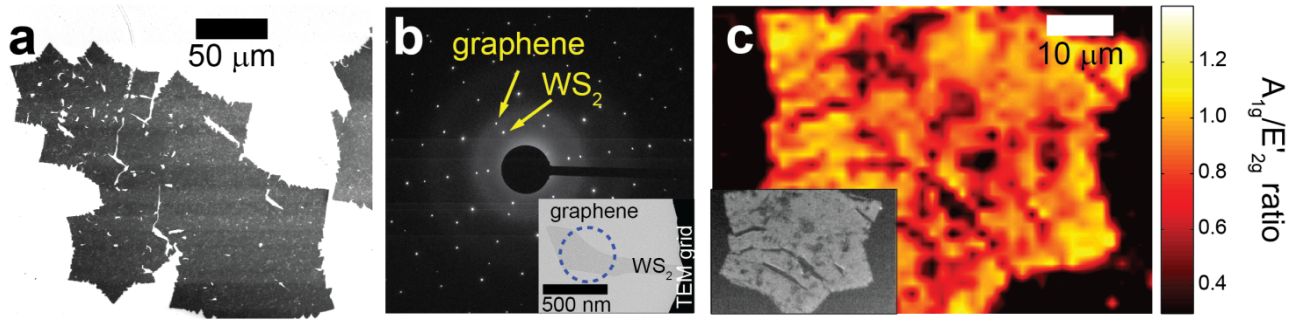

**Figure 2.** (a) SEM image of CVD graphene crystals transferred on quartz and fully covered by $WS_2$. (b) Selected area electron microdiffraction pattern of the heterostack collected from the area enclosed by the dotted circle in the bright field image reported in the inset. (c) Raman map of the ratio between the intensities of $A_{1g}$ and $E_{2g}^1$ peaks. Inset: optical image of the selected flake used for Raman.

We now turn to the analysis of $WS_2$ synthesized on graphene, an ideal substrate to transduce the collected quantum information into a photocurrent. Fig. 2(a) is a SEM micrograph of CVD graphene single-crystals after growth of a continuous $WS_2$ film. The crystallinity of the $WS_2$ layer is evaluated by analyzing a partial growth via TEM imaging and selected area electron microdiffraction (details in Supporting Information). As shown in Figure 2(b), the partial process leads to the formation of $WS_2$ single-crystals with an epitaxial relation to the graphene substrate. The single-crystal domains appear to be several hundreds of nanometers in size. The Raman map reported in panel (c) indicates that most of the graphene grains are covered by bilayer $WS_2$ (i.e., $I(A_{1g}/E_{2g}^1) \sim 1$)[23] (see also Supporting Information). Local monolayer inclusions are observed especially in proximity and within transfer-generated tears of graphene (which are visible in the optical image in panel (c)).

*3.3 Self-organized photo-emitting/photo-conductive patterns of $WS_2$ on epitaxial graphene.*

As PL is quenched in $WS_2$/graphene stacks because of enhanced electron transfer to gapless graphene[29], one might wonder whether the synthesized $WS_2$ films still presents a robust polarization conservation. To gain some insight into this, we also consider epitaxial graphene on SiC(0001)[30] as a perfect platform to investigate the optical properties of the synthesized $WS_2$ crystals. On a typical epitaxial monolayer graphene sample, monolayer areas alternate with zerolayer areas following the atomic terraces of the SiC surface[31]. Zerolayer, or buffer layer, graphene is the first carbon-rich layer forming on top of SiC and, although the hexagonal lattice is retained, 30% of the constituting atoms are covalently bonded to the substrate[32]. Local breaking of the $sp^2$ hybridization disrupts the semi-metallic nature of the layer and hence its quenching properties. In the SEM micrograph reported in Fig. 3(a), one can appreciate triangular $WS_2$ grains obtained with a partial growth covering monolayer graphene (light gray contrast) and zerolayer graphene (mid gray contrast). Indeed, the PL intensity from a sample covered with $WS_2$ bilayers (see thickness assessment in Supporting Information) is altered along lines parallel to the terrace directions (see panel (c)). A measurable photoemission is detected only in correspondence of $WS_2$/zerolayer stacks. Analogously to the h-BN substrate, an elongated 2D histogram of PL polarization and intensity is observed and the polarization conservation is again peaked at 72%. The use of epitaxial graphene on SiC as a substrate for $WS_2$ synthesis presents a scalable bottom-up approach for patterning ribbons, which can either emit or transfer the charge to the graphene underneath. The outstandingly high mobility and free path length in graphene potentially allows to transport the quantum information from the absorbed light via spin-polarized electrons.

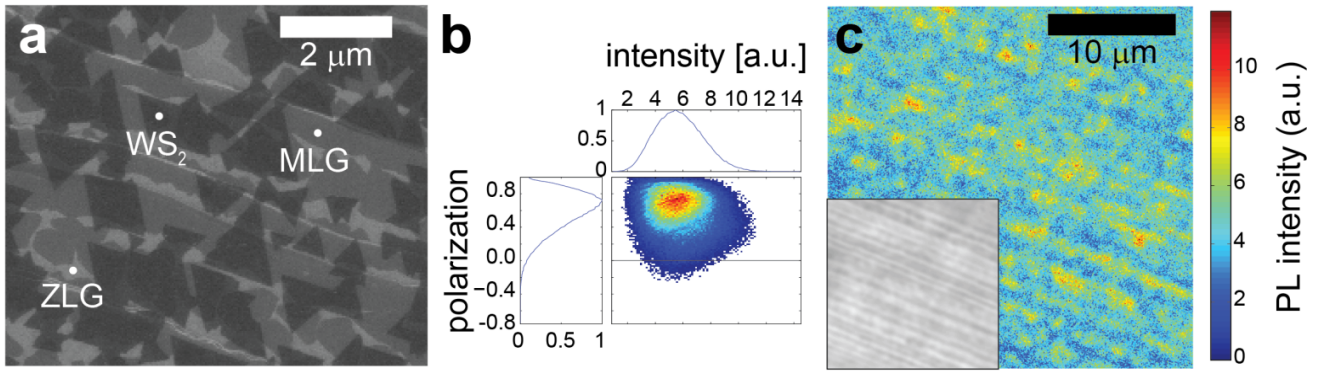

**Figure 3.** (a) SEM image of WS$_2$ on epi-graphene. Brighter areas are attributed to buffer layer, intermediate grey to graphene and darker triangles to WS$_2$ crystals. (b) PL polarization vs. PL intensity 2D histogram. The polarization peaks at ~70%. (c) PL intensity map taken from the area shown in the inset. The emission in stronger where WS$_2$ is on top of buffer layer regions.

## 4. Conclusion

In conclusion, we have demonstrated the CVD synthesis of continuous atomically-thick WS$_2$ films on 2D substrates. The synthesized films display remarkable polarization conservation at room temperature, as high as 74%. We show that by adopting epitaxial graphene on SiC as growth substrate, one can define in a bottom-up fashion photoemitting and photoconducting ribbons. The scalable synthesis and design on 2D substrates of WS$_2$ films with outstanding optical properties is instrumental in the development of novel all-2D quantum optoelectronic and valleytronic devices.

**Acknowledgments**

The authors would like to thank Fabio Beltram from NEST-Scuola Normale Superiore, Giovanni Signore from CNI@NEST and Iwan Moorels from Istituto Italiano di Tecnologia for fruitful discussion. The research leading to these results has received funding from the European Union Seventh Framework Program under grant agreement no. 604391 Graphene Flagship.

## References

[1] M. Chhowalla, H. S. Shin, G. Eda, L.-J. Li, K. P. Loh, H. Zhang, *Nature Chemistry* **2013**, 5, 263.
[2] H.-P. Komsa, A. V. Krasheninnikov, *Physical Review B* **2013**, 88, 085318.
[3] Q. H. Wang, K. Kalantar-Zadeh, A. Kis, J. N. Coleman, M. S. Strano, *Nature Nanotechnology* **2012**, 7, 699.
[4] H. R. Gutiérrez, N. Perea-López, A. L. Elías, A. Berkdemir, B. Wang, R. Lv, F. López-Urías, V. H. Crespi, H. Terrones, M. Terrones, *Nano Letters* **2012**, 13, 3447.
[5] Z. Ye, T. Cao, K. O'Brien, H. Zhu, X. Yin, Y. Wang, S. G. Louie, X. Zhang, *Nature* **2014**, 513, 214.
[6] B. Zhu, X. Chen, X. Cui, *Scientific Reports* **2015**, 5, 9218.
[7] B. Zhu, H. Zeng, J. Dai, Z. Gong, X. Cui, *Proceedings of the National Academy of Sciences* **2014**, 111, 11606.
[8] P. K. Nayak, F.-C. Lin, C.-H. Yeh, J.-S. Huang, P. W. Chiu, *Nanoscale* **2016,** 6035.
[9] S. Bae, H. Kim, Y. Lee, X. Xu, J.-S. Park, Y. Zheng, J. Balakrishnan, T. Lei, H. R. Kim, Y. I. Song, *Nature Nanotechnology* **2010**, 5, 574.
[10] N. Mishra, V. Miseikis, D. Convertino, M. Gemmi, V. Piazza, C. Coletti, *Carbon* **2016**, 96, 497.
[11] Y. Rong, Y. Fan, A. L. Koh, A. W. Robertson, K. He, S. Wang, H. Tan, R. Sinclair, J. H. Warner, *Nanoscale* **2014**, 6, 12096.
[12] Q. Fu, W. Wang, L. Yang, J. Huang, J. Zhang, B. Xiang, *RSC Advances* **2015**, 5, 15795.
[13] G. V. Bianco, M. Losurdo, M. M. Giangregorio, A. Sacchetti, P. Prete, N. Lovergine, P. Capezzuto, G. Bruno, *RSC Advances* **2015**, 5, 98700.
[14] M. Okada, T. Sawazaki, K. Watanabe, T. Taniguch, H. Hibino, H. Shinohara, R. Kitaura, *ACS Nano* **2014**, 8, 8273.
[15] Y. Zhang, Y. Zhang, Q. Ji, J. Ju, H. Yuan, J. Shi, T. Gao, D. Ma, M. Liu, Y. Chen, *ACS Nano* **2013**, 7, 8963.
[16] A. L. Elias, N. Perea-López, A. Castro-Beltrán, A. Berkdemir, R. Lv, S. Feng, A. D. Long, T. Hayashi, Y. A. Kim, M. Endo, *ACS Nano* 2013, 7, 5235.
[17] V. Miseikis, D. Convertino, N. Mishra, M. Gemmi, T. Mashoff, S. Heun, N. Haghighian, F. Bisio, M. Canepa, V. Piazza, *2D Materials* **2015**, 2, 014006.
[18] F. Bianco, D. Perenzoni, D. Convertino, S. L. De Bonis, D. Spirito, M. Perenzoni, C. Coletti, M. S. Vitiello, A. Tredicucci, *Applied Physics Letters* **2015**, 107, 131104.
[19] A. S. Mayorov, R. V. Gorbachev, S. V. Morozov, L. Britnell, R. Jalil, L. A. Ponomarenko, P. Blake, K. S. Novoselov, K. Watanabe, T. Taniguchi, *Nano Letters* **2011**, 11, 2396.
[20] C. Cong, J. Shang, X. Wu, B. Cao, N. Peimyoo, C. Qiu, L. Sun, T. Yu, *Advanced Optical Materials* **2014**, 2, 131.
[21] A. Berkdemir, H. R. Gutiérrez, A. R. Botello-Méndez, N. Perea-López, A. L. Elías, C.-I. Chia, B. Wang, V. H. Crespi, F. López-Urías, J.-C. Charlier, *Scientific Reports* **2013**, 3, 1755.
[22] H. Li, Q. Zhang, C. C. R. Yap, B. K. Tay, T. H. T. Edwin, A. Olivier, D. Baillargeat, *Advanced Functional Materials* **2012**, 22, 1385.

[23]   A. Thangaraja, S. M. Shinde, G. Kalita, M. Tanemura, *Applied Physics Letters* **2016**, 108, 053104.
[24]   N. Peimyoo, J. Shang, W. Yang, Y. Wang, C. Cong, T. Yu, *Nano Research* **2015**, 8, 1210.
[25]   W. Zhao, Z. Ghorannevis, K. K. Amara, J. R. Pang, M. Toh, X. Zhang, C. Kloc, P. H. Tan, G. Eda, *Nanoscale* **2013**, 5, 9677.
[26]   Y. Gao, Z. Liu, D.-M. Sun, L. Huang, L.-P. Ma, L.-C. Yin, T. Ma, Z. Zhang, X.-L. Ma, L.-M. Peng, *Nature Communications* **2015**, 6, 8569.
[27]   L. Yuan, L. Huang, *Nanoscale* **2015**, 7, 7402.
[28]   T. C. Berkelbach, M. S. Hybertsen, D. R. Reichman, Physical Review B 2013, 88, 045318.
[29]   A. Kasry, A. A. Ardakani, G. S. Tulevski, B. Menges, M. Copel, L. Vyklicky, *The Journal of Physical Chemistry C* **2012**, 116, 2858.
[30]   C. Berger, Z. Song, T. Li, X. Li, A. Y. Ogbazghi, R. Feng, Z. Dai, A. N. Marchenkov, E. H. Conrad, P. N. First, *The Journal of Physical Chemistry B* **2004**, 108, 19912.
[31]   C. Coletti, C. L. Frewin, S. E. Saddow, M. Hetzel, C. Virojanadara, U. Starke, *Applied Physics Letters* **2007**, 91, 61914.
[32]   S. Goler, C. Coletti, V. Piazza, P. Pingue, F. Colangelo, V. Pellegrini, K. V. Emtsev, S. Forti, U. Starke, F. Beltram, *Carbon* **2013**, 51, 249.

# Supporting Information

# Scalable synthesis of WS$_2$ on graphene and h-BN: an all-2D platform for light-matter transduction

*Antonio Rossi,*[†‡] *Holger Büch,*[†] *Carmine Di Rienzo,*[†‡] *Vaidotas Miseikis,*[†] *Domenica Convertino,*[†‡] *Ameer Al-Temimy,*[†] *Valerio Voliani,*[†] *Mauro Gemmi,*[†] *Vincenzo Piazza*[†] *and Camilla Coletti*[§†*]

[†]Center for Nanotechnology Innovation @NEST, Istituto Italiano di Tecnologia, Piazza S. Silvestro 12, 56127 Pisa, Italy

[‡]NEST-Scuola Normale Superiore, Piazza San Silvestro 12, 56127 Pisa, Italy

[§]Graphene Labs, Istituto Italiano di Tecnologia, Via Morego 30, 16163 Genova, Italy

*corresponding author: camilla.coletti@iit.it

**Additional Methods**

*Graphene growth, h-BN transfer, SiC and quartz preparation:* The substrates used in this work were: (i) h-BN exfoliated flakes on quartz; (ii) monolayer graphene obtained via chemical vapor deposition and transferred onto quartz; (iii) epitaxial monolayer graphene on SiC(0001). Details about their preparation follow.

(i) h-BN flakes were obtained via mechanical exfoliation on quartz substrates by scotch tape method[1]. Before exfoliation, the quartz substrate was cleaned in acetone and isopropanol, and in O$_2$ plasma (5 min, 80W). After the exfoliation, the sample was cleaned again in acetone and isopropanol, and treated with oxygen to remove possible scotch tape residue.

(ii) Growth of large single-crystals of monolayer graphene was performed on Cu foil as reported in Ref.[2]. Graphene was then transferred onto quartz substrate via dry transfer.

(iii) Monolayer graphene was grown on nominally on-axis 6H–SiC(0001). The SiC substrate was first subjected to HF treatment to remove native oxides, and then hydrogen etched at ~ 1250°C to prepare an atomically-flat 6H–SiC(0001) surface[3]. Graphene was subsequently obtained by annealing the atomically-flat sample for several minutes in argon atmosphere of 780 mbar at about 1700 K in a resistively heated cold-wall reactor (BM, Aixtron)[4].

*Raman spectroscopy and SEM imaging:* Raman characterization was performed using a standard Renishaw inVia system equipped with a 532 nm green laser and 100x objective lens. The laser spot size was ~1$\mu$m and the accumulation time was 1s. Scanning Electron Microscopy (SEM) imaging was performed at 5 keV using a Zeiss Merlin microscope, equipped with a field emission gun.

*Trasmission electron microscopy:* Transmission electron microscopy was carried out on a Ziess Libra 120 transmission electron microscope operating at 120 kV and equipped with an in-column Omega filter for energy filtered imaging. Both electron diffraction patterns and bright field images were recorded energy filtered, with a 20 eV slit centered on the zero loss peak. Electron diffraction patterns were collected with a parallel beam in micro-diffraction mode, by selecting the diffracting area through the appropriate condenser aperture. For sample preparation, graphene was transferred on a gold grid[2]. Subsequently, $WS_2$ was directly grown onto this system.

**Additional Raman data**

Raman analysis is a powerful tool to reveal the number of layer of WS$_2$ and also to test the quality of the substrate such as graphene. Figure S1(a) shows typical Raman spectra of WS$_2$ on h-BN and epitaxial graphene, highlighting the 310 cm$^{-1}$ peak. The Raman spectra collected for mono- and bilayer WS$_2$ regions grown on CVD graphene are shown in panel (b). Figure S1(c) shows the spectra measured for CVD graphene before and after WS$_2$ growth, demonstrating that graphene is not damaged by the growth process. Figure S1(d) shows typical Raman spectra taken from the zerolayer and monolayer graphene regions. The 2D peak is significantly lowered in the latter.

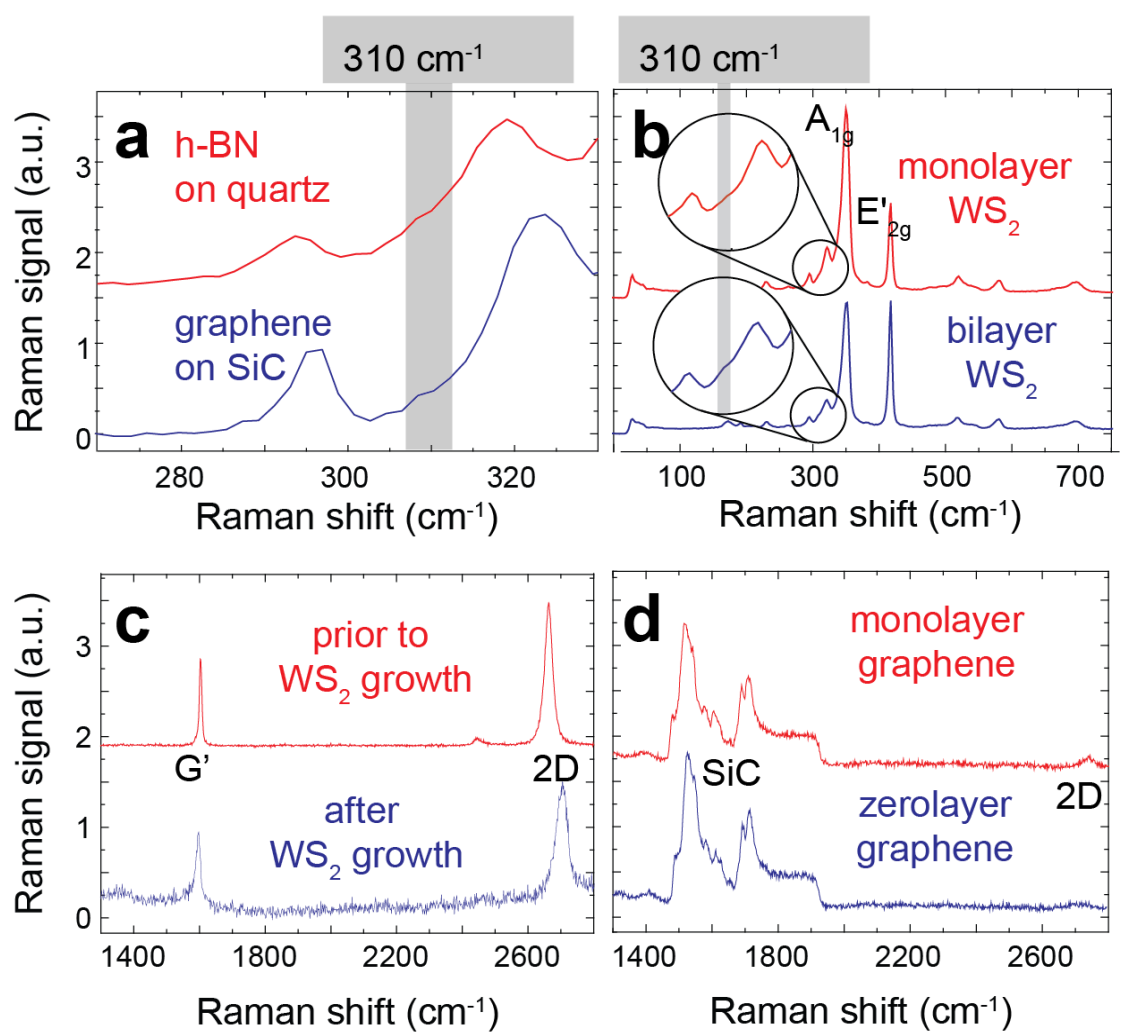

**Figure S1 : Raman Analysis:** (a) Raman peak in the 310 cm$^{-1}$ range of WS$_2$ on h-BN and Graphene on Silicon Carbide. (b) Raman spectra of WS$_2$ on CVD graphene taken in area mostly bilayer (blue) and monolayer (red). (c) Graphene Raman spectrum before (red) and after the growth (blue), indicating that graphene is intact also after the process. (d) Raman spectrum of epitaxial graphene (red) and buffer layer (blue), taken within the same substrate.

**AFM analysis of WS$_2$ partial growth on epitaxial graphene**

The epitaxial graphene topography was investigated via AFM before and after WS$_2$ growth. The phase image reported in Figure S2(a) shows typical monolayer (darker) and zerolayer (brighter) graphene regions along the atomic terraces of SiC (Figure S2(b))[5,6]. Panels (c) and (d) are a representative AFM micrograph and the relative line profile taken after WS$_2$ growth and showing the mostly bilayer coverage of our sample.

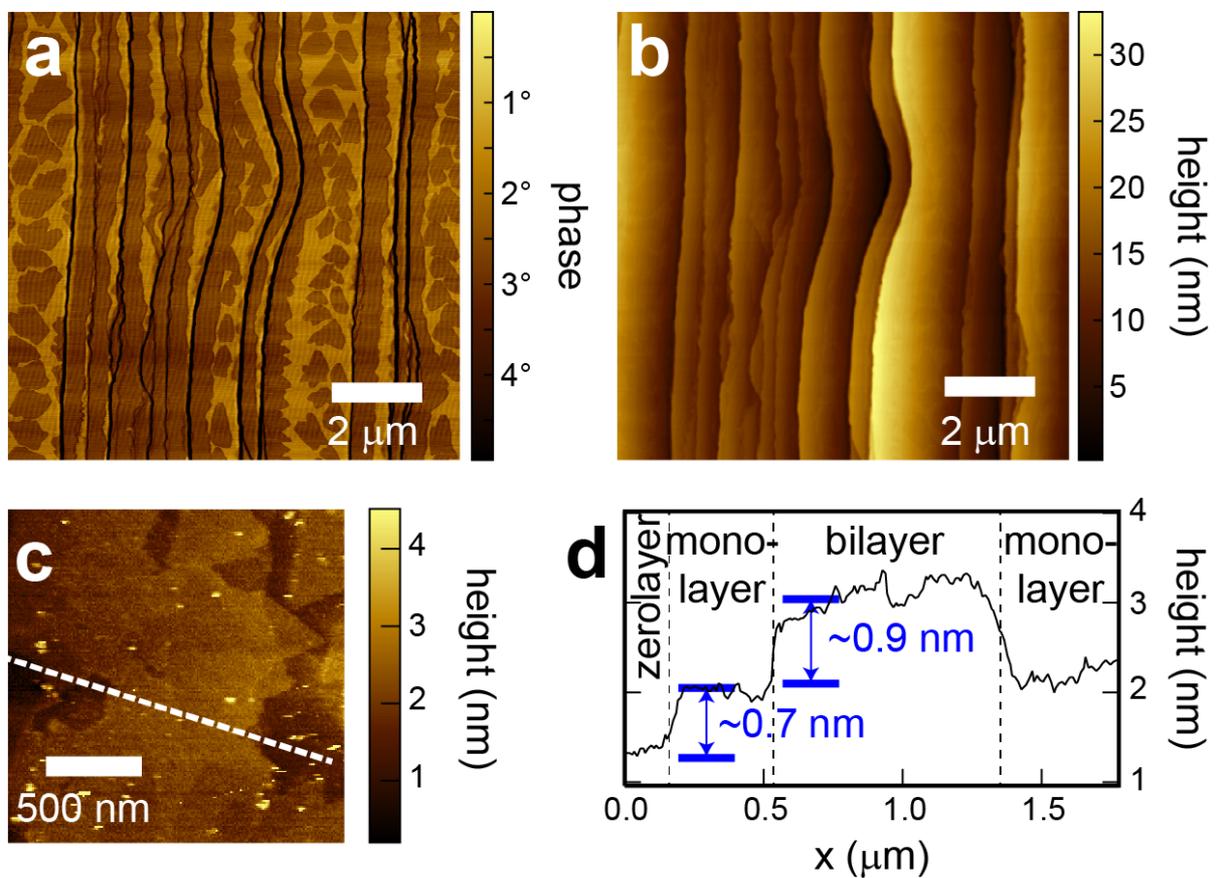

**Figure S2 : Atomic Force Microscopy Scan:** phase (a) and height (b) scan of graphene on silicon carbide before the growth. The two colors in the phase panel indicate graphene and buffer layer respectively. (c) Height scan of the same sample after WS$_2$ growth. (d) Profile of the highlighted line, showing monolayer and bilayer regions.

1. N. Mishra, V. Miseikis, D. Convertino, M. Gemmi, V. Piazza, C. Coletti, *Carbon* **2016,** 96, 497-502.
2. V. Miseikis, D. Convertino, N. Mishra, M. Gemmi, T. Mashoff, S. Heun, N. Haghighian, F. Bisio, M. Canepa, V. Piazza, *2D Materials* **2015,** 2, (1), 014006.
3. C. Coletti, C. L. Frewin, S. E. Saddow, M. Hetzel, C. Virojanadara, U. Starke, *Applied Physics Letters* **2007**, 91, 61914.
4. Bianco, F.; Perenzoni, D.; Convertino, D.; De Bonis, S. L.; Spirito, D.; Perenzoni, M.; Coletti, C.; Vitiello, M. S.; Tredicucci, A. *Applied Physics Letters* **2015,** 107, (13), 131104.
5. S. Goler, C. Coletti, V. Piazza, P. Pingue, F. Colangelo, V. Pellegrini, K.V. Emtsev, S. Forti, U. Starke, F. Beltram, *Carbon* **2013,** 51, 249-254
6. S. Frewin, C. Coletti, C. Riedl, U. Starke, S.E. Saddow, *Mater. Sci. Forum* **2009** 615–617, 589.